\documentclass{article}[12pt]
\usepackage{graphicx}

\begin{document}

\begin{center}
\Large{\bf Estimate of the Three-Loop $\overline {MS}$ Contribution to 
$\sigma(W_L^+ W_L^- \rightarrow Z_L Z_L)$} 
\end{center}

\bigskip

\begin{center}
F. A. Chishtie and V. Elias\\
Department of Applied Mathematics\\
University of Western Ontario\\
London, Ontario N6A 5B7\\
Canada
\end{center}

\smallskip

\begin{abstract}
The three-loop contribution to the $\overline{MS}$ single-Higgs-doublet standard-model 
cross-section $\sigma(W_L^+ W_L^- \rightarrow Z_L Z_L)$ at  $s = (5M_H)^2$ is estimated 
via least-squares matching of the asymptotic Pad\'e-approximant prediction of the next 
order term, a procedure that has been previously applied to QCD corrections to correlation 
functions and decay amplitudes. In contrast to these prior applications, the expansion 
parameter for the $ W_L^+ W_L^- \rightarrow Z_L Z_L$ process is the non-asymptotically-free
quartic scalar-field coupling of the standard model, suggesting that the least-squares matching be
performed over the ``infrared'' $\mu^2 \leq s$  region of the scale parameter. All three coefficients of
logarithms within the three-loop term obtained by such matching are found to be within 6.6\%
relative error of their true values, as determined via renormalization-group methods. 
Surprisingly, almost identical results are obtained by performing the least squares
matching over the $\mu^2 \geq s$ region.    
\end{abstract}

Within standard-model single-Higgs-doublet electroweak physics, the cross-section for the
scattering of two longitudinal W's into two longitudinal Z's at very high energy takes the 
form

\begin{equation}
\sigma[s, L(\mu), g(\mu)] = \frac{8\pi^3}{9s} H[s, L (\mu), g (\mu) ],
\end{equation}
where the scale-sensitive portion of the cross-section [1],
\begin{eqnarray}
H[s, L(\mu), g(\mu)] & = & g^2 (\mu) \left\{ 1 + [-4 L (\mu) - 10.0] g(\mu) \right. \nonumber \\
& + & \left[ 12 L^2 (\mu) + 68.667 L (\mu) + \left( 93.553 + \frac{2}{3} ln \left( \frac{s}{M_H^2} 
\right) \right) \right] g^2 (\mu) \nonumber \\
& + & \left. \left[ c_3 L^3 (\mu) + c_2 L^2 (\mu) + c_1 L (\mu) + c_0 \right] g^3 (\mu) + ... \right\},
\end{eqnarray}
depends on the renormalization scale $\mu$ explicitly through the logarithm
\begin{equation}
L(\mu) \equiv ln (\mu^2 / s)
\end{equation}
and implicitly through the $\overline{MS}$ quartic scalar-field coupling 
\begin{equation}
g(\mu) = 6 \; \lambda_{\overline{MS}} (\mu) / 16 \pi^2.
\end{equation}

The three-loop coefficients $\{ c_0, c_1, c_2, c_3 \}$ in (2) are presently unknown. 
The factor of $M_H$ appearing explicitly in the two-loop term calculated in [1] is a 
scale-independent pole mass; the coefficients $\{ c_0, c_1, c_2, c_3 \}$ can also exhibit 
dependence on this mass without acquiring additional $\mu$-dependence.

In this note, we utilise asymptotic Pad\'e-approximant methods to predict these four
coefficients.  Such methods have already been applied to predicting next-order terms within the
renormalization group functions of QCD [2,3,4], supersymmetric QCD [2,5], and massive scalar
field theory [3,4,6], as well as next-order QCD corrections to scalar and vector current correlation
functions [4,7] and various decay processes [8,9,10].   In all of these applications, the Pad\'e-
estimation procedures are tested (with surprising success) against either those higher-order terms
already known from explicit calculation, such as renormalization group functions in scalar field
theories [11,12], QCD [13], and supersymmetric QCD [14], or against those coefficients of logarithms [such as $ \{ c_1, c_2, c_3 \}$ 
in (2)] which can be extracted via renormalization group methods [7,8,9,10].  Below, we shall apply
the latter testing procedure to estimates of the next-order terms $\{ c_1, c_2, c_3 \}$ in
the $\overline {MS}$ cross-section (1).

Given a perturbative series of the form $1 + R_1 g + R_2 g^2 + R_3 g^3 + ...$ where $R_3$ is 
not known, as is the case in (2), asymptotic Pad\'e-approximant methods can be employed to 
show that [4]
\begin{equation}
R_3 \cong \frac{2 R_2^3}{R_1^3 + R_1 R_2} .
\end{equation}
This result is, of course, contingent upon the field-theoretical series exhibiting appropriate
asymptotic behaviour.  Its derivation (explicitly presented in [8]) follows from the $O(1/N)$ 
error anticipated from an $[N|1]$ Pad\'e-approximant prediction of $R_{N+2}$ [15], a 
semi-empirical behaviour which is seen to characterise a number of field-theoretical 
applications even when $N$ is small.

For the case of the series (2), however, $R_1$ is linear in $L$ and $R_2$ is quadratic in $L$. 
Consequently, the prediction (5) for $R_3$ corresponds to a rational function of $L$ 
incompatible with the degree-3 polynomial in $L$ anticipated from (2).  Clearly, a procedure 
is required by which predictions for the polynomial coefficients $\{ c_0, c_1, c_2, c_3 \}$ in (2) can 
be extracted from (5).  In past applications where the same problem arises [7,8,9,10], one 
method employed is a least-squares matching of (5) to the form
$R_3 = c_0 + c_1 L(\mu) + c_2 L^2(\mu) + c_3 L^3(\mu)$ over the full perturbative domain of $\mu$. 
For QCD calculations this domain is ultraviolet; {\it e.g.} in estimating three-loop QCD 
corrections to $B \rightarrow X_u \ell^- \overline{\nu}_\ell$ the matching is over the
ultraviolet domain $\mu \geq m_b(\mu)$ [8]. For the expression (2), in which the perturbative 
expansion parameter is the non-asymptotically-free quartic scalar coupling $\lambda_{\overline{MS}}(\mu)$,
the appropriate domain for such a least-squares matching is {\it infrared}. 

Thus, to obtain predicted values for $\{ c_0, c_1, c_2, c_3 \}$ for a given choice of $M_H$, we 
choose a least-squares matching over the region $\mu^2 \leq s$, or alternatively $0 < w \leq 1$, 
where $w \equiv \mu^2/s$ is the argument of the logarithm (3). From (2) and (5), this matching 
is achieved by minimizing the function

\newpage

\begin{eqnarray}
&& \chi^2(c_0, c_1, c_2, c_3) \equiv \nonumber \\
&& \int_{w_{min}}^1 dw \left[ \frac{2 R_2^3 (w)}{R_1^3 (w) + R_1 (w) R_2
(w)} - c_0 - c_1 ln (w) - c_2 ln^2 (w) - c_3 ln^3 (w) \right]^2 \; \; \; \;
\end{eqnarray}
with respect to $c_0, c_1, c_2, c_3$, where $R_1(w)$ and $R_2(w)$ are explicitly given in (2):
\begin{equation}
R_1 (w) = -4 ln (w) - 10.0,
\end{equation}
\begin{equation}
R_2 (w) = 12 ln^2 (w) + 68.667 ln (w) + \left( 93.553 + \frac{2}{3}
ln \left( \frac{s}{M_H^2} \right) \right).
\end{equation}
The lower bound of integration $w_{min}$ in (6) would ordinarily be zero
to encompass the full $\mu^2 \leq s$ range.  However, we are compelled
to consider a nonzero value of $w_{min}$ in order to avoid any integrand
poles, as discussed below.  The expressions (1) and (2) are stated in ref. [1]
to be accurate (within single-digit percent errors) only in the high-energy
limit $\sqrt{s} \stackrel{>}{_\sim} 5M_H$. Although the projected linear and quadratic 
dependence of $c_1$ and $c_0$ on $ln(s/M_H^2)$ could, in principle, be extracted via Pad\'e 
methods,\footnote{In ref. [10], the polynomial dependence of three-loop order terms in
$H \rightarrow gg$ on the logarithm of the pole-mass ratio $M_H / M_t$
is similarly extracted.} the relatively small coefficient of this
logarithm in the previous-order term (8) necessarily implies a similar insensitivity to this 
logarithm in Pad\'e estimates of next-order terms.  Consequently, we restrict our analysis here 
to the $s = (5M_H)^2$ kinematic boundary of applicability for (1) and (2).  
With this choice, the integrand of (6) acquires singularities at 0.0552,
0.0821, and 0.0896.  Consequently, we choose $w_{min} = 0.09$ to include
virtually all of the integrable infrared region, and we find that
\begin{eqnarray}
\chi^2 (c_0, c_1, c_2, c_3) \nonumber \\
& = & 248033 + 813.779c_0 - 332.772c_1 + 244.574c_2 - 242.664c_3 \nonumber \\
& + & 0.91c_0^2 - 1.38657c_0 c_1 + 1.72946 c_0 c_2 \nonumber \\
& - & 2.67527c_0 c_3 + 0.864732 c_1^2 - 2.67527c_1 c_2 + 4.64965 c_1 c_3 \nonumber \\
& + & 2.32482 c_2^2 - 8.67669 c_2 c_3 + 8.48631 c_3^2
\end{eqnarray}
By then optimizing (9) with respect to $c_0, c_1, c_2, c_3$, we obtain the following Pad\'e 
predictions for these coefficients:
\begin{equation}
c_0^{Pad\acute{e}} = -896, c_1^{Pad\acute{e}} = -889, c_2^{Pad\acute{e}}
= -288, c_3^{Pad\acute{e}} = -30.5 .
\end{equation}

As noted earlier, the true values of the coefficients $c_1, c_2, c_3$ can be extracted via 
the scale-invariance of the physical cross-section (1):
\begin{equation}
O = \mu^2 \frac{dH}{d\mu^2} [s, L(\mu), g(\mu)] = \frac{\partial
H}{\partial L} + \beta (g) \frac{\partial H}{\partial g},
\end{equation}
where [11]
\begin{equation}
\beta(g) = 2g^2 - \frac{13}{3} g^3 + 27.803 g^4 + ... \; .
\end{equation}
One can verify that the known terms in (2) satisfy the renormalization-group equation (11) to 
$O(g^3)$ and $O(g^4)$, as is evident from the series expansions
\begin{equation}
\frac{\partial H}{\partial L} = -4g^3 + (24 L + 68.667) g^4 + (3c_3 L^2
+ 2c_2 L + c_1) g^5 + ... \; ,
\end{equation}
\begin{eqnarray}
\beta(g) \frac{\partial H}{\partial g} & = & 4g^3 + (-24 L - 68.667) g^4 +
(96 L^2 + 601.33 L + 934.028 \nonumber \\
& + & 5.333 ln (s/M_H^2)) g^5 + ... \; .
\end{eqnarray}
We find upon incorporating $O(g^5)$ terms of (13) and (14) into the right-hand side of (11) that
\begin{eqnarray}
c_1 & = & -934.028 - 5.333 ln (s/M_H^2) 
\begin{array}{c}{ } \\ \longrightarrow \\ _{s = 25 M_H^2}
\end{array}
-951.2,\nonumber \\
c_2 & = & -300.67, \; c_3 = -32.
\end{eqnarray}
The  Pad\'e predictions (10) for $c_1, c_2, c_3$ are respectively seen to be within relative 
errors of 6.6\%, 4.3\%, and 4.7\% of their true values (15).  

Curiously, the accuracy of these 
Pad\'e results does not appear to be contingent upon the matching being performed over the ``infrared'' 
$\mu^2 < s = 25 \; M_H^2$ range, as motivated by the non-asymptotically free character of the scalar field 
coupling $g(\mu)$.  Indeed, a potential drawback of fitting over the
$\mu^2 \leq s$ (hence, $w \leq 1$) region, as in (6), is the negativity
of $ln(w)$ over the entire range of integration.  Since the $c_i$
ultimately obtained in (10) are all same-sign (negative), cancellations
necessarily occur between successive $c_k ln^k (w)$ terms in the
integrand of (6) in the best-fit region of $c_k$ parameter-space.  To
address this issue, we have also performed a fit of the 
Pad\'e-prediction (5) to the third-subleading order of (2) 
over the entire $\mu^2 \geq s$ ({\it i.e.} $w \geq 1$) region in which $ln(w)$ is {\it positive}.  This
entails integration of the integrand of (6) with appropriately modified bounds of integration
to encompass the ultraviolet region:
\begin{equation}
\int_{w_{min}}^1 dw [. . .]^2 \rightarrow \int_1^\infty dw [. . .]^2 .
\end{equation}  
We then find that    
\begin{eqnarray}
\chi^2 (c_0, c_1, c_2, c_3) & = & 1.45665 \cdot 10^7 + 5095.28 c_0 + 10294.6 c_1 \nonumber \\
& + & 35519.6 c_2 + 167166 c_3 + c_0^2 + 2 c_0 c_1 \nonumber \\
& + & 4 c_0 c_2 + 12 c_0 c_3 + 2 c_1^2 + 12 c_1 c_2 + 48 c_1 c_3 \nonumber \\
& + & 24 c_2^2 + 240 c_2 c_3 + 720 c_3^2,
\end{eqnarray}
which upon optimization, yields values for $c_k [ c_0 = -896, \;
c_1 = -889, c_2 = -289 \; c_3 = -30.9]$ that are virtually the
same as those listed in (10). The small relative errors characterising our
Pad\'e estimates of $c_1, c_2, c_3$ are comparable 
to those characterising Pad\'e  estimates of renormalization-group accessible coefficients 
within next-order QCD corrections to other processes [7,8,9,10], and suggest similar accuracy 
in the estimated value of the renormalization-group {\it inaccessible} three-loop coefficient $c_0$ in (10).

We therefore conclude that Pad\'e-approximant predictions of the next order contribution 
to $WW \rightarrow ZZ$ at very high energies appear to be consistent and reliable.
It should also be noted that higher order $\beta$-function 
terms associated with the evolution of the quartic scalar-field coupling constant (4) are 
themselves accurately 
predicted by the same asymptotic Pad\'e-approximant methods employed above for $\sigma(WW \rightarrow ZZ)$. 
If we express the $\beta$-function (12) in the form
\begin{equation}
\beta(g) = 2g^2(1+R_1 g + R_2g^2 + R_3 g^3 + ...); \; \; R_1 = -13/6, \; R_2 = 13.915,
\end{equation}
we predict via (5) that $R_3 = -133.6$, or alternatively, that the predicted next term in the 
series (12) is $-267.2g^5$. This is quite close to $-266.495g^5$, the true calculated value [11] 
of the next-order $\beta$-function contribution.  Similarly close agreement between the somewhat 
more complicated asymptotic Pad\'e-approximant prediction and the explicit calculation of the $O(g^6)$ 
contribution to this $\beta$-function is demonstrated in ref. [6].

	Thus, the results presented above are an example of how Pad\'e estimation 
procedures can anticipate next-order contributions whose exact values are obtainable only 
by lengthy calculation. For the particular process in question, the distinction between two- 
and three-loop order results is seen to be unimportant unless the mass of the (Salam-Weinberg) 
Higgs field mediating the scattering process is very large.  This is illustrated in Figures 1-3,
which compare two loop and three loop expressions for the scale sensitive portion (2) of the 
$WW \rightarrow ZZ$ cross section (1) for Higgs-field masses of 200, 400, and 600 GeV, respectively.  Only for 
the largest of these three choices is an appreciable difference anticipated between two- and three-loop order predictions 
for the cross section.  

\clearpage
\begin{figure}[hbt]
\centering
\includegraphics[scale=0.6, angle=270]{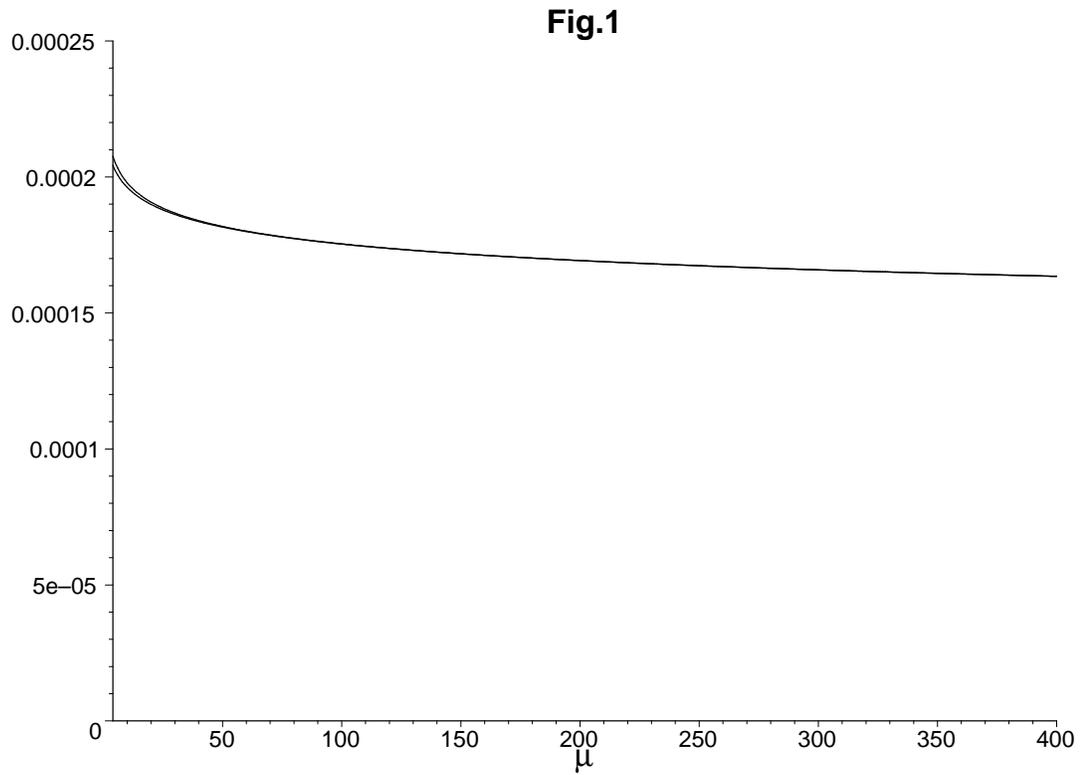}
\caption{The two-loop (bottom curve) and predicted three-loop (top curve) expressions for
the scale-sensitive portion $H[s=(5 M_H)^2$,$L(\mu)$,$g(\mu)]$ of the cross-section (1)
are plotted for Higgs mass $M_H=200$ GeV.
}
\label{fig1}
\end{figure}

\clearpage
\begin{figure}[hbt]
\centering
\includegraphics[scale=0.6, angle=270]{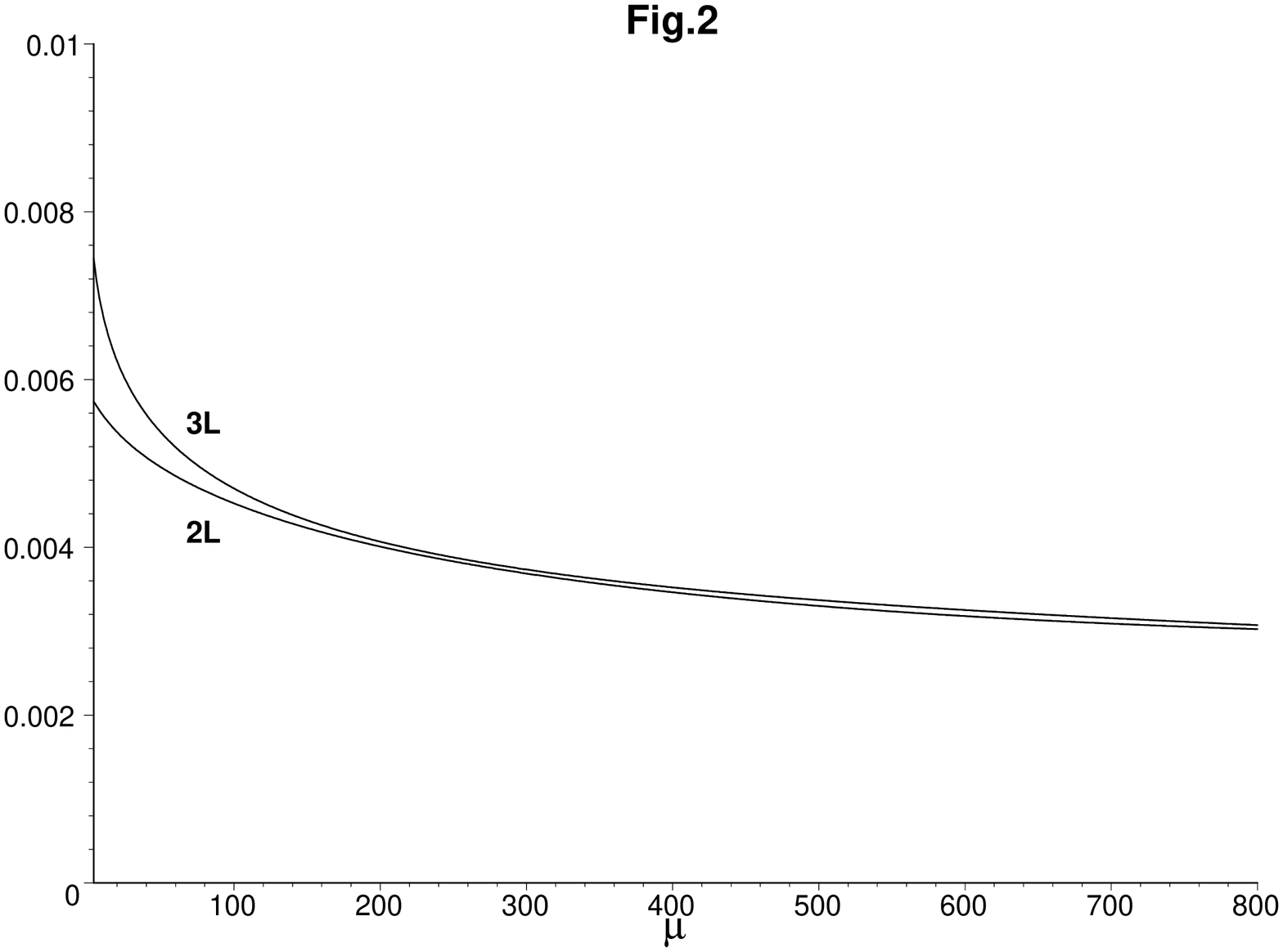}
\caption{Comparison of two-loop (2L) and three-loop (3L) expressions, as in Figure 1, but with $M_H = 
400$ GeV.
}
\label{fig2}
\end{figure}

\clearpage
\begin{figure}[hbt]
\centering
\includegraphics[scale=0.6, angle=270]{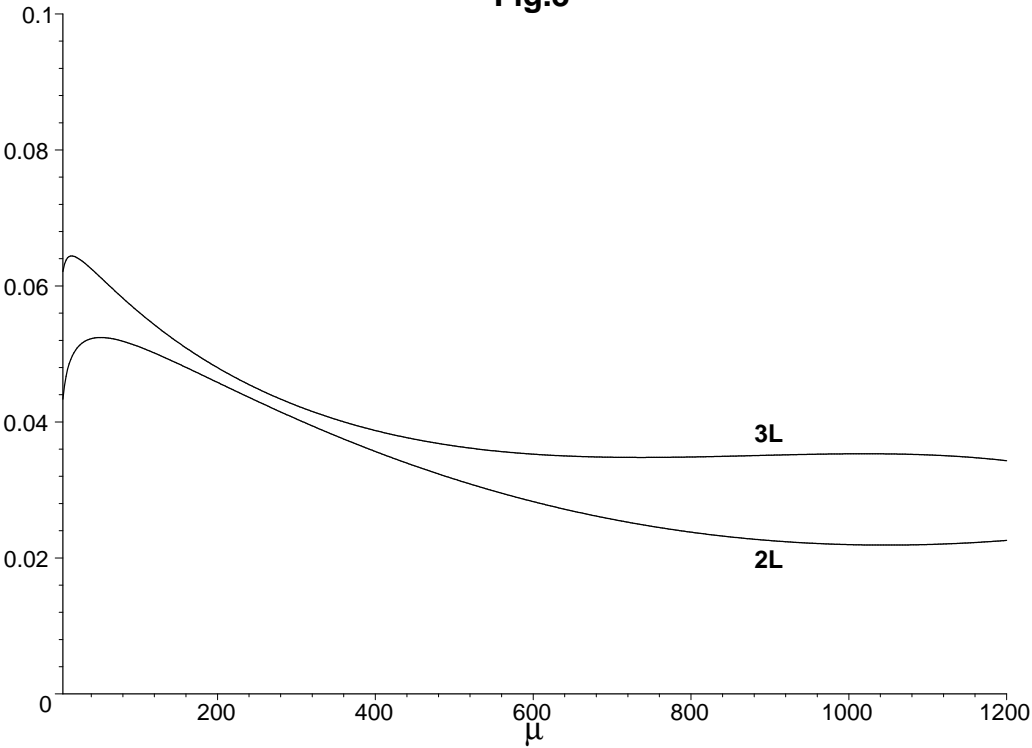}
\caption{Comparison of two-loop (2L) and three-loop (3L) expressions, as in Figure 1, but with $M_H = 
600$ GeV.
}
\label{fig3}
\end{figure}

\clearpage

\section*{Acknowledgment}

VE is grateful for research support from the Natural Sciences and Engineering Research Council of
Canada.

\section*{References}

1.  U. Nierste and K. Riesselmann, Phys. Rev. D 53 (1996) 6638.\\
2.  J. Ellis, I. Jack, D.R.T. Jones, M. Karliner, and M. A. Samuel,
Phys. Rev. \indent D 57 (1998) 2665.\\
3.  J. Ellis, M. Karliner, and M. A. Samuel, Phys. Lett. B 400 (1997)
176.\\
4.  V. Elias, T. G. Steele, F. Chishtie, R. Migneron, and K. Sprague,
Phys. Rev. \indent D 58 (1998) 116007.\\
5.  I. Jack, D.R. T. Jones, and M. A. Samuel, Phys. Lett. B 407 (1997)
143.\\
6.  F. Chishtie, V. Elias, and T. G. Steele, Phys. Lett B 446 (1999)
267.\\
7.  F. Chishtie, V. Elias, and T. G. Steele, Phys. Rev. D 59 (1999)
105013.\\
8.  M. R. Ahmady, F. A. Chishtie, V. Elias, and T. G. Steele, Phys.
Lett. B \indent 479 (2000) 201.\\
9.  F. A. Chishtie, V. Elias, and T. G. Steele, J. Phys. G 26 (2000)
93.\\
10.  F. A. Chishtie, V. Elias, and T. G. Steele, J. Phys. G 26 (2000)
1239.\\
11.  H. Kleinert, J. Neu, V. Schulte-Frohlinde, K. G. Chetyrkin, and S.
A. Larin, \indent Phys. Lett. B 272 (1991) 39 and (Erratum) B 319 (1993) 545.\\
12.  B. Kastening, Phys. Rev. D 57 (1998) 3567.\\
13.  J. A. M. Vermaseren, S. A. Larin, and T. Van Ritbergen, Phys. Lett. B\\
\indent 405 (1997) 327; K. G. Chetyrkin, Phys. Lett. B 404 (1997) 161.\\
14.  I. Jack, D. R. T. Jones, and A. Pickering, Phys. Lett. B 435 (1998) 61.\\
15.  M. A. Samuel, J. Ellis, and M. Karliner, Phys. Rev. Lett. 74 (1995) 4380;\\
\indent J. Ellis, E. Gardi,
M. Karliner, and M. A. Samuel, Phys. Lett. B 366 (1996) \\
\indent 268 and Phys. Rev. D 54 (1996) 6986; S. J. Brodsky,
J. Ellis, E. Gardi, M.\\
\indent Karliner, and M. A. Samuel, Phys. Rev. D 56 (1997) 6980.
\end{document}